\documentclass[%
 reprint,
superscriptaddress,
showpacs,
 amsmath,amssymb,
 aps,
pre,
]{revtex4-1}

\usepackage{graphicx}
\usepackage{color}

\begin{document}


\title{Dynamics of two populations of phase oscillators with different frequency distributions}

\author{Yu Terada}
\email{y-terada@acs.i.kyoto-u.ac.jp}
\affiliation{Graduate School of Informatics, Kyoto University, Kyoto 606-8501, Japan}
\author{Toshio Aoyagi}
\affiliation{Graduate School of Informatics, Kyoto University, Kyoto 606-8501, Japan}


\date{\today}

\begin{abstract}
A large variety of rhythms are observed in nature.
Rhythms such as electroencephalogram signals in the brain can often be regarded as interacting.
In this study, we investigate the dynamical properties of rhythmic systems in two populations of phase oscillators with different frequency distributions.
We assume that the average frequency ratio between two populations closely approximates some small integer.
Most importantly, we adopt a specific coupling function derived from phase reduction theory.
Under some additional assumptions, the system of two populations of coupled phase oscillators reduces to a low-dimensional system in the continuum limit.
Consequently, we find chimera states in which clustering and incoherent states coexist.
Finally, we confirm consistent behaviors of the derived low-dimensional model and the original model.

\end{abstract}

\maketitle


\section{\label{sec:level1}Introduction}

Rhythm plays a crucial role in many aspects of physics, biology and chemistry \cite{strogatz2003sync,winfree2001geometry}.
To study the rhythmic phenomena quantitatively, the phase oscillator model  has been widely used since the prominent studies had been done \cite{winfree1967biological,kuramoto1975self,ermentrout1991multiple}. 
The Kuramoto model undergoes a transition from nonsynchronous to synchronous as the coupling strength increases.
This model can also be solved exactly with the Lorentzian natural frequency distribution \cite{kuramoto1984chemical}.
Although this discovery has triggered many subsequent works, many unsolved, related models are considered important in real world situations \cite{acebron2005kuramoto}.
One of such important models is a system composed of multiple heterogeneous populations of phase oscillators.
In fact, some real systems appear to have a hierarchical structure of multiple populations.
For example, multiple neuronal modules in the brain seem to organize into structural networks \cite{bullmore2009complex,buzsaki2009rhythms}.
To understand the dynamical prosperities of such systems,  we need to theoretically investigate a model with multiple populations of oscillators.
As a first step, we here investigate a two-population system of phase oscillators.

It is plausible that the characteristics of two populations are generally different and in several situations this property seems to play an important functional role.
For example, the synchrony of different neuronal populations in the brain is positively correlated with the success of human tasking \cite{buzsaki2009rhythms}.
In real systems such as electroencephalograms, the average frequencies often largely differ across populations.
Considering such a coupled system of two populations with different average frequency distributions, we can theoretically derive the correct form of the coupling function by the averaging method \cite{sanders2007averaging} in the phase reduction method.
This derivation is essential if average frequencies between two populations are related by an integer ratio such as $k:1$ (where $k$ is an integer). This condition is called the resonant condition. \cite{luck2011dynamics,komarov2013dynamics}.
However, resonance has not been considered in most previous studies.
Therefore, we will examine multifrequency oscillator systems by applying a specific coupling function.

Moreover, multiple populations of phase oscillators exhibit interesting properties \cite{okuda1991mutual,abrams2008solvable,komarov2011effects,komarov2013dynamics}.
One of the most remarkable phenomena is the formation of chimera states in which synchronous and asynchronous states coexist \cite{montbrio2004synchronization,abrams2008solvable,laing2009chimera,laing2012disorder,martens2010bistable,martens2010chimeras,pazo2014low,laing2010chimeras}.
This phenomenon was theoretically discovered by Kuramoto \textit{et al.} \cite{kuramoto2002coexistence} and named by Abrams \textit{et al.} \cite{abrams2004chimera}.
Later, the properties of chimera states were experimentally investigated \cite{tinsley2012chimera,hagerstrom2012experimental,nkomo2013chimera}.
Recent theoretical works have explored chimera states in more general situations \cite{motter2010nonlinear,gu2013spiral,panaggio2013chimera,singh2011chimera,omelchenko2013nonlocal}.
However, in most studies on chimera states in multiple phase oscillator populations, the natural frequencies of the oscillators are assumed to be evenly distributed across the populations.
As the next stage, we should theoretically examine heterogeneous frequency distributions across the populations.
To this end, we treat a general situation in which the oscillator populations have different average frequencies.
We focus on the resonant case with an integer frequency ratio.
Here, we must derive the appropriate type of phase coupling function under the corresponding resonance condition.
We show that such a system develops chimera states under some conditions and investigate the properties of these states.

Recently Ott and Antonsen proposed a remarkable ansatz that reduces an infinite system of coupled phase oscillators in the continuum limit to a low-dimensional system \cite{ott2008low,ott2009long,ott2011comment}.
This ansatz has been applied to a wide range of applications and has yielded many fruitful results  \cite{martens2009exact,lee2009large,montbrio2011shear,skardal2012hierarchical,kloumann2014phase,tanaka2014solvable,mirollo2012asymptotic}.
Komarov and Pikovsky considered the resonant interactions among more than two oscillator communities and applied the Ott-Antonsen ansatz to a simple resonant case \cite{komarov2013dynamics}.
In this study, we consider two populations of phase oscillators in the more general resonant case $k:1$, where $k$ is not $\pm1$.
However, because the populations interact through the resonant type coupling function, we cannot straightforwardly reduce the system to low-dimensional equations using the Ott-Antonsen ansatz.
To proceed with the analysis, we augment the Ott-Antonsen ansatz with the additional assumptions of Skardal \textit{et al.}.
Consequently, our system reduces to a three-dimensional system of ordinary differential equations.

The remainder of this paper is structured as follows.
Section \ref{sec:model} describes a coupled system of two populations of phase oscillators.
In Sec. \ref{sec:reduction}, we reduce this system to a low-dimensional system.
The emergent dynamical states such as the clustered chimera states are investigated in Sec. \ref{sec:results}.
In Sec. \ref{sec:comparison}, we numerically confirm that our results hold without the additional solvability assumption.
Our method is extended to more general resonant conditions in Sec. \ref{sec:general}.
The paper concludes with a summary in Sec. \ref{sec:discussion}.

\section{Phase reduction and Model equations}\label{sec:model}

As a preliminary, we discuss two interacting oscillators.
The oscillators evolve by the following equations:
\begin{align}
\frac{d\boldsymbol{x}_1}{dt} = \boldsymbol{f}_1\left(\boldsymbol{x}_1\right) + \epsilon \boldsymbol{g}_1\left(\boldsymbol{x}_1,\boldsymbol{x}_2\right), \label{eq: original_oscillators1}\\
\frac{d\boldsymbol{x}_2}{dt} = \boldsymbol{f}_2\left(\boldsymbol{x}_2\right) + \epsilon \boldsymbol{g}_2\left(\boldsymbol{x}_2,\boldsymbol{x}_1\right), \label{eq: original_oscillators2}
\end{align}
where $\boldsymbol{x}_1$ and $\boldsymbol{x}_2$ are $n$-dimensional state vectors, $\boldsymbol{f}_1$ and $\boldsymbol{f}_2$ represent the intrinsic dynamics of the oscillators,  and $\boldsymbol{g}_1$ and $\boldsymbol{g}_2$ are the interaction terms between the oscillators.
We further suppose $\lvert\epsilon\rvert\ll1$, and that the oscillators have non-perturbed limit cycles (when $\epsilon=0$).
The periods of oscillators 1 and 2 are $2\pi/\omega_1$ and $2\pi/\omega_2$ respectively, where $\omega_1$ and $\omega_2$ are the respective natural frequencies of the oscillators.
The frequencies almost satisfy the resonant relation $k:1$, where the natural frequency of the fast oscillator $\omega_1$ is approximately $k$ times the frequency of the slow oscillator $\omega_2$:
\begin{align}
\omega_1\simeq k\omega_2.\label{eq:almost_resonant_relation}
\end{align}
If the frequencies satisfy Eq. (\ref{eq:almost_resonant_relation}), the resonant coupling function can be derived by phase reduction.

We first introduce the original phase variables $\theta_1$ and $\theta_2$ such that $d\theta_1/dt=\omega_1 , d\theta_2/dt=\omega_2$ near the limit cycle orbits $\boldsymbol{x}_{1,0}(t)$ and $\boldsymbol{x}_{2,0}(t)$ in the absence of the perturbations.
Applying phase reduction, the dynamics of the phase variables $\theta_1$ and $\theta_2$ are determined from Eqs. (\ref{eq: original_oscillators1}) and (\ref{eq: original_oscillators2}) as
\begin{align}
\frac{d\theta_1}{dt} = k\omega+\epsilon\boldsymbol{Z}_1\left(\theta_1\right)\cdot\boldsymbol{g}_{12}\left(\theta_1,\theta_2\right), \label{eq: original_phase_dynamics1}\\
\frac{d\theta_2}{dt} = \omega+\epsilon\boldsymbol{Z}_2\left(\theta_2\right)\cdot\boldsymbol{g}_{21}\left(\theta_2,\theta_1\right),\label{eq: original_phase_dynamics2}
\end{align}
where
\begin{align}
\boldsymbol{Z}_1\left(\theta_1\right) &= \boldsymbol{\nabla}_{\boldsymbol{x}_1}\theta_1\left(\boldsymbol{x}_1\right)|_{\boldsymbol{x}_1=\boldsymbol{x}_{1,0}\left(\theta_1\right)},\notag\\
\boldsymbol{Z}_2\left(\theta_2\right) &= \boldsymbol{\nabla}_{\boldsymbol{x}_2}\theta_2\left(\boldsymbol{x}_2\right)|_{\boldsymbol{x}_2=\boldsymbol{x}_{2,0}\left(\theta_1\right)}.\notag
\end{align}
To separate the slow dynamics from Eqs.(\ref{eq: original_phase_dynamics1}) and (\ref{eq: original_phase_dynamics2}), we define slow phase variables $\psi_1$ and $\psi_2$ as $\theta_1=2\omega t+\psi_1$ and $\theta_2=\omega t+\psi_2$ respectively.
The dynamics of the slow phase variables are described by
\begin{align}
\frac{d\psi_1}{dt} = &\epsilon\boldsymbol{Z}_1\left(\psi_1+k\omega t\right)\cdot\boldsymbol{g}_{12}\left(\psi_1+k\omega t,\psi_2+\omega t\right), \label{eq: slow_phase_dynamics1}\\
\frac{d\psi_2}{dt} &= \epsilon\boldsymbol{Z}_2\left(\psi_2+\omega\right)\cdot\boldsymbol{g}_{21}\left(\psi_2+\omega t,\psi_1+k\omega t\right).\label{eq: slow_phase_dynamics2}
\end{align}
Averaging the RHS in Eq. (\ref{eq: slow_phase_dynamics1}) over the period of the slow oscillator $2\pi/\omega$, we obtain
\begin{align}
\frac{d\psi_1}{dt} &= \epsilon\frac{\omega}{2\pi}\int^{2\pi/\omega}_{0}dt\boldsymbol{Z}_1\left(\psi_1+k\omega t\right)\cdot\boldsymbol{g}_{12}\left(\psi_1+k\omega t,\psi_2+\omega t\right) \notag\\
&=  \frac{\epsilon}{2\pi}\int^{2\pi}_{0}d\Theta\boldsymbol{Z}_1\left(\psi_1+k\Theta\right)\cdot\boldsymbol{g}_{12}\left(\psi_1+k\Theta,\psi_2+\Theta\right)\notag\\
&=  \frac{\epsilon}{2\pi}\int^{2\pi}_{0}d\Theta\boldsymbol{Z}_1\left(\psi_1-k\psi_2+k\Theta\right)\cdot\notag\\
&\qquad\qquad\qquad\qquad\boldsymbol{g}_{12}\left(\psi_1-k\psi_2+k\Theta,\Theta\right),\label{eq: slow_averaging_first}
\end{align}
where $\psi_1$ and $\psi_2$ are constants independent of $t$  during one period.
When evaluating the integral, we express the RHS of Eq. (\ref{eq: slow_averaging_first}) as a function of $\psi_1-k\psi_2$; namely $\Gamma_{12}\left(\psi_1-k\psi_2\right)$.
Performing similar operations the RHS of Eq. (\ref{eq: slow_phase_dynamics2}), we obtain the pair of averaged equations:
\begin{align}
\frac{d\psi_1}{dt} = \epsilon\Gamma_{12}\left(\psi_1-k\psi_2\right), \label{eq: averaged_slow_phase_dynamics1}\\
\frac{d\psi_2}{dt} = \epsilon\Gamma_{21}\left(k\psi_2-\psi_1\right),\label{eq: averaged_slow_phase_dynamics2}
\end{align}
where
\begin{align}
\Gamma_{12}\left(\psi_1-k\psi_2\right) &= \frac{\omega}{2\pi}\int^{2\pi/\omega}_{0}dt \boldsymbol{Z}_1\left(\psi_1+k\omega t\right)\cdot\notag\\
&\boldsymbol{g}_{12}\left(\psi_1+k\omega t,\psi_2+\omega t\right),\notag\\
\Gamma_{21}\left(k\psi_2-\psi_1\right) &= \frac{\omega}{2\pi}\int^{2\pi/\omega}_{0}dt \boldsymbol{Z}_2\left(\psi_2+\omega t\right)\cdot\notag\\
&\boldsymbol{g}_{21}\left(\psi_2+\omega t,\psi_1+k\omega t\right).\notag
\end{align}
This approximation is valid to order $\epsilon$.
Consequently, we obtain the evolutionary equations of the averaged phase variables $\theta_1$ and $\theta_2$:
\begin{align}
\frac{d\theta_1}{dt} = k\omega + \epsilon \Gamma_{12}\left(\theta_1-k\theta_2\right), \\
\frac{d\theta_2}{dt} = \omega + \epsilon \Gamma_{21}\left(k\theta_2-\theta_1\right).
\end{align}
We emphasize that the coupling functions $\Gamma_i$ ($i$=1,2) depend on $\theta_1-k\theta_2$ (or $k\theta_2-\theta_1$) when the natural frequencies of the oscillators satisfy the resonant relation $k:1$.
Generalizing this to the $m:n$ case, we can state that $\Gamma_i$ ($i$=1,2) depends on $n\theta_1-m\theta_2$ (or $m\theta_2-n\theta_1$).

\begin{figure}
\includegraphics[scale=0.55]{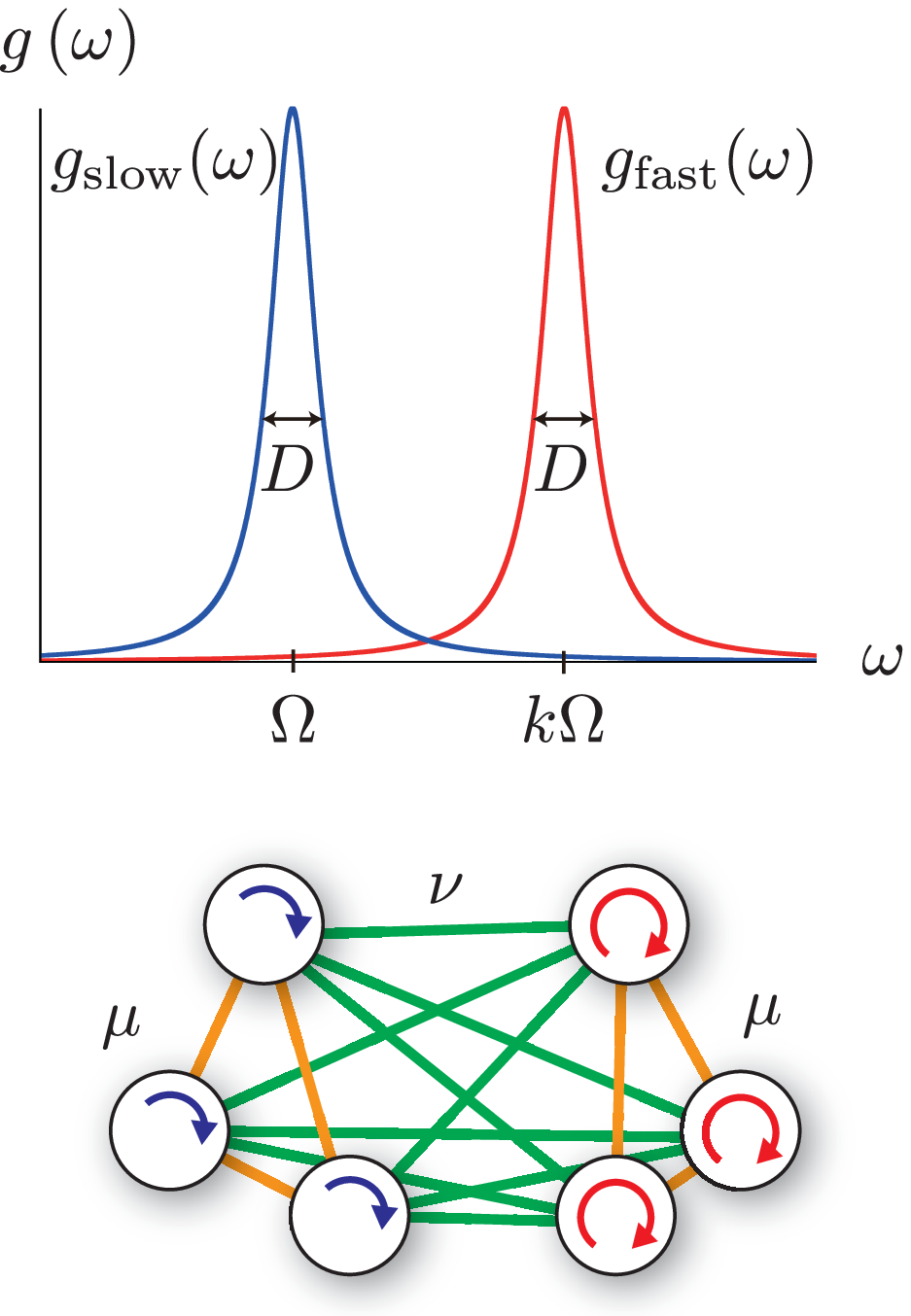}
 \caption{\label{fig freq_dist}Top: Natural frequency distributions of the phase oscillators in the two populations whose mean frequencies satisfy the resonant relation $k:1$.\\ Bottom: The coupling strengths is $\mu=(1+A)/2$ within the same population and $\nu=(1-A)/2$ between the two populations.}
\end{figure}

Our discussion of resonant interactions is now extended to populations of phase oscillators.
We consider two populations of phase oscillators with different average frequencies.
We assume that both populations have inherent Lorentzian (Cauchy) distributions of natural frequencies.
The mean frequency ratio between the two populations is $2:1$ (i.e., $k=2$ in Fig. \ref{fig freq_dist}):
\begin{align}
g_{\text{fast}}\left(\omega\right) = \frac{D}{\pi}\frac{1}{\left(\omega-2\Omega\right)^2+D^2},\label{eq:lorentz_2omega} \\
g_{\text{slow}}\left(\omega\right) = \frac{D}{\pi}\frac{1}{\left(\omega-\Omega\right)^2+D^2},\label{eq:lorentz_omega}
\end{align}
where $D$ is the common width of the distributions and $\Omega$ is the mean of the distribution in the slow population.
As usual, we assume dominance of the first term in the Fourier series of the coupling function, and take $H\left(\theta\right)=\sin\left(\theta+\alpha\right)$ as in the Kuramoto-Sakaguchi model \cite{sakaguchi1986soluble}, where $\alpha$ is the phase lag parameter.
We thus consider the following model:
\begin{align}
\displaystyle\frac{d\theta^{\text{fast}}_i}{dt} &= 
\begin{aligned}[t]
\omega^{\text{fast}}_i +& \displaystyle\frac{\mu}{N}\displaystyle\sum^{N}_{j=1}\sin\left(\theta^{\text{fast}}_j-\theta^{\text{fast}}_i-\alpha\right)\\
&+ \displaystyle\frac{\nu}{N}\displaystyle\sum^{N}_{j=1}\sin\left(2\theta^{\text{slow}}_j-\theta^{\text{fast}}_i-\alpha\right),\label{eq:original1} \\
\end{aligned}\\
\displaystyle\frac{d\theta^{\text{slow}}_i}{dt} &=
\begin{aligned}[t]
\omega^{\text{slow}}_i +& \displaystyle\frac{\mu}{N}\displaystyle\sum^{N}_{j=1}\sin\left(\theta^{\text{slow}}_j-\theta^{\text{slow}}_i-\alpha\right)\\
&+ \displaystyle\frac{\nu}{N}\displaystyle\sum^{N}_{j=1}\sin\left(\theta^{\text{fast}}_j-2\theta^{\text{slow}}_i-\alpha\right), \label{eq:original2}
\end{aligned}
\end{align}
where $\theta^{\text{fast/slow}}_i$ is the phase of oscillator $i$ ($i=1,\cdots,N$) in the fast/slow population.
Incorporating the phase reduction concept, we now consider the interactions between the oscillators.
For this purpose, we introduce the parameter $A$ and define $\mu=\left(1+A\right)/2, \nu=\left(1-A\right)/2$ as in  \cite{abrams2008solvable}.
In this setting, $\mu+\nu=1$.
The parameter $A$ controls the ratio of the strengths of the interactions within each population and between the populations.
When $0<A<1$, the interactions are stronger within the populations than across the populations; conversely, when $-1<A<0$ the interactions across the populations dominate. 
In this study we fix the phase lag parameter $\alpha=\pi/2-0.05$ as in  \cite{laing2012disorder}.
Under this condition, chimera states will appear, and a variety of dynamics with a reasonably general coverage are expected \cite{abrams2008solvable}.
The general $k$ case will be straightforwardly extended from the $k=2$ case in Sec. \ref{sec:general}.

\section{Reduction to Low-dimensional dynamics}\label{sec:reduction}

Using the Ott-Antonsen ansatz \cite{ott2008low,ott2009long,ott2011comment}, we will attempt to reduce the system represented by Eqs. \eqref{eq:lorentz_2omega}-\eqref{eq:original2} to a low-dimensional system in the limit $N\to\infty$.
To handle the high-order Kuramoto model, we use a modified version of the Ott-Antonsen ansatz used by Skardal \textit{et al.} \cite{skardal2011cluster}.
However, the modified ansatz alone does not reduce the original dynamics to a low-dimensional system.
To overcome this difficulty, we employ one additional assumption in which we replace $\sin\left(\theta^{\text{slow}}_j-\theta^{\text{slow}}_i-\alpha\right)$ with $\sin\left(2\theta^{\text{slow}}_j-2\theta^{\text{slow}}_i-\alpha\right)$  in Eq. \eqref{eq:original2}.
Equation \eqref{eq:original2} becomes
\begin{align}
\displaystyle\frac{d\theta^{\text{slow}}_i}{dt} &=
\begin{aligned}[t]
\omega^{\text{slow}}_i +& \displaystyle\frac{\mu}{N}\displaystyle\sum^{N}_{j=1}\sin\left(2\theta^{\text{slow}}_j-2\theta^{\text{slow}}_i-\alpha\right)\\
&+ \displaystyle\frac{\nu}{N}\displaystyle\sum^{N}_{j=1}\sin\left(\theta^{\text{fast}}_j-2\theta^{\text{slow}}_i-\alpha\right).\label{eq:modified2}
\end{aligned}
\end{align}
The dynamics can be reduced by applying the Ott-Antonsen ansatz to \eqref{eq:original1} and the modified model Eq. \eqref{eq:modified2}.
In general, however, it is unlikely that the interaction terms among the slow oscillators contain no first Fourier mode.
In Sec. \ref{sec:comparison}, we will check the validity of the modified model under general conditions.

We now consider the continuum limit $N\to\infty$ in our modified model.
The probability density functions (PDFs) of the fast and slow populations in the continuum limit are denoted as $f_{\text{fast}}(\theta,\omega,t)$ and $f_{\text{slow}}(\theta,\omega,t)$ respectively, where  $f_j(\theta,\omega,t)d\theta d\omega dt$ is the fraction of oscillators with phase between $\theta$ and $\theta+d\theta$ and natural frequency between $\omega$ and $\omega+d\omega$ at time $t$ in population $j(=\text{fast,slow})$.
The common order parameter for the fast population is given by
\begin{align}
\displaystyle z_{\text{fast}}(t)
	&=\lim_{N\to\infty}\displaystyle\frac{1}{N}\sum^{N}_{j=1}e^{i\theta^{\text{fast}}_j}\notag\\
	&= \int^{\infty}_{-\infty}d\omega\int^{2\pi}_{0}d\theta f_{\text{fast}}\left(\theta,\omega,t\right)e^{i\theta}.\label{eq:fast_order_parameter}
\end{align}
For the slow population, we define the Daido order parameter \cite{daido1992order} as
\begin{align}
\displaystyle z_{\text{slow}}(t)
	&=\lim_{N\to\infty}\displaystyle\frac{1}{N}\sum^{N}_{j=1}e^{2i\theta^{\text{slow}}_j}\notag\\
	&= \int^{\infty}_{-\infty}d\omega\int^{2\pi}_{0}d\theta f_{\text{slow}}\left(\theta,\omega,t\right)e^{2i\theta}.\label{eq:slow_order_parameter}
\end{align}
Following the Ott-Antonsen ansatz \cite{ott2008low,ott2009long,ott2011comment} and its variant \cite{skardal2011cluster}, we expand the PDFs as two Fourier series:
\begin{align}
f_{\text{fast}}(\theta,\omega,t) &= \frac{g_{\text{fast}}(\omega)}{2\pi}\left[1+\sum^{\infty}_{n=1}\left(a(\omega,t)^ne^{in\theta}+\text{c.c.}\right)\right], \label{eq: fast_density}\\
f_{\text{slow}}(\theta,\omega,t) &= \frac{g_{\text{slow}}(\omega)}{2\pi}\left[1+\sum^{\infty}_{m=1}\left(b(\omega,t)^{m}e^{2im\theta}+\text{c.c.}\right)\right],\label{eq: slow_density}
\end{align}
where c.c. stands for complex conjugate.
The ansatz requires the conditions, $\lvert a\left(\omega,t\right)\rvert<1$ and $\lvert b\left(\omega,t\right)\rvert<1$ for the convergence of Eqs. \eqref{eq: fast_density} and \eqref{eq: slow_density}.
To conserve the total number of oscillators in each population, the following continuity equations should be satisfied:
\begin{align}
\frac{\partial f_{j}}{\partial t}+\frac{\partial}{\partial\theta_{j}}\left(f_{j}\dot{\theta}_{j}\right) = 0\,\,\,\left(j=\text{fast},\text{slow}\right),\notag
\end{align} 
Substituting Eqs. \eqref{eq: fast_density} and \eqref{eq: slow_density} into these continuity equations, we obtain the ordinal differential equations for $a$ and $b$:
\begin{align}
\frac{\partial a}{\partial t}+i\omega a&+\frac{\mu}{2}\left(z_{\text{fast}}a^2e^{-i\alpha}-\bar{z}_{\text{fast}}e^{i\alpha}\right)\notag\\
&+\frac{\nu}{2}\left(z_{\text{slow}}a^2e^{-i\alpha}-\bar{z}_{\text{slow}}e^{i\alpha}\right) = 0,\label{eq:a}\\
\frac{1}{2}\frac{\partial b}{\partial t}+i\omega b&+\frac{\mu}{2}\left(z_{\text{slow}}b^2e^{-i\alpha}-\bar{z}_{\text{slow}}e^{i\alpha}\right)\notag\\
&+\frac{\nu}{2}\left(z_{\text{fast}}b^2e^{-i\alpha}-\bar{z}_{\text{fast}}e^{i\alpha}\right) = 0.\label{eq:b}
\end{align}
From Eqs. \eqref{eq:fast_order_parameter} and \eqref{eq: fast_density}, we immediately find that
\begin{align}
z_{\text{fast}}\left(t\right) = \int^{\infty}_{-\infty}d\omega g_{\text{fast}}\left(\omega\right)\bar{a}\left(\omega,t\right).
\end{align}
Given that the natural frequency distribution is Lorentzian Eq. \eqref{eq:lorentz_2omega}, we can write
\begin{align}
z_{\text{fast}}\left(t\right) = \frac{1}{2\pi i}\int^{\infty}_{-\infty}d\omega &\left(\frac{1}{\omega-2\Omega-iD}-\frac{1}{\omega-2\Omega+iD}\right)\notag\\
&\times\bar{a}\left(\omega,t\right).\label{eq:z_int}
\end{align}
Following \cite{ott2008low} we assume that $\bar{a}\left(\omega,t\right)$ is analytic in $\text{Im}\,\omega>0$.
From the complex conjugate of Eq. \eqref{eq:a}, we know that $\partial\bar{a}/\partial t\sim -\left(\text{Im}\,\omega\right)\bar{a}$ as $\text{Im}\,\omega\to\infty$.
As $\lvert\bar{a}\left(\omega,t\right)\rvert<1$, we also know that $\bar{a}\left(\omega,t\right)\to0$ as $\text{Im}\,\omega\to\infty$.
Integrating the RHS of Eq. \eqref{eq:z_int}  over the upper semicircular contour in the complex plain, we obtain
\begin{align}
z_{\text{fast}}\left(t\right) = \bar{a}\left(2\Omega+iD,t\right).\notag
\end{align}
A similar analysis gives 
\begin{align}
z_{\text{slow}}\left(t\right) = \bar{b}\left(\Omega+iD,t\right).\notag
\end{align}
Substituting $\omega=2\Omega+iD$ and $\omega=\Omega+iD$ in the complex conjugate of Eqs. \eqref{eq:a} and \eqref{eq:b}, respectively, we obtain the dynamics of the complex order parameters of the two populations:
\begin{align}
\frac{dz_{\text{fast}}}{dt} &=
\begin{aligned}[t]
&\left(-D+2\Omega i\right)z_{\text{fast}} + \displaystyle\frac{e^{-i\alpha}}{2}\left(\mu z_{\text{fast}}+\nu z_{\text{slow}}\right) \\
&-\displaystyle\frac{e^{i\alpha}}{2}\left(\mu\bar{z}_{\text{fast}}+\nu\bar{z}_{\text{slow}}\right)z_{\text{fast}}^2,\label{eq:complex_dyn1}
\end{aligned}\\
\frac{dz_{\text{slow}}}{dt} &=
\begin{aligned}[t]
&\left(-2D+2\Omega i\right)z_{\text{slow}}+e^{-i\alpha}\left(\mu z_{\text{slow}}+\nu z_{\text{fast}}\right) \\
&-e^{i\alpha}\left(\mu\bar{z}_{\text{slow}}+\nu\bar{z}_{\text{fast}}\right)z_{\text{slow}}^2.\label{eq:complex_dyn2}
\end{aligned}
\end{align}
We now rewrite Eqs. \eqref{eq:complex_dyn1} and \eqref{eq:complex_dyn2} in terms of the polar coordinates $z_{\text{fast}}=r_{\text{fast}}e^{-i\phi_{\text{fast}}}$ and $z_{\text{slow}}=r_{\text{slow}}e^{-i\phi_{\text{slow}}}$ and their phase difference $\phi=\phi_{\text{fast}}-\phi_{\text{slow}}$.
The resulting system comprises three ordinary differential equations with three degrees of freedom $(r_{\text{fast}},r_{\text{slow}},\phi)$:
\begin{align}
\frac{dr_{\text{fast}}}{dt} &=
 \begin{aligned}[t]
&-Dr_{\text{fast}} \\
&+\displaystyle\frac{1-r_{\text{fast}}^2}{2}\left(\mu r_{\text{fast}}\cos\alpha+\displaystyle\nu r_{\text{slow}}\cos\left(\phi-\alpha\right)\right), \label{eq: three_dynamics1} \\
\end{aligned}\\
\frac{dr_{\text{slow}}}{dt} &=
 \begin{aligned}[t]
&-2Dr_{\text{slow}} \\
&+\displaystyle\frac{1-r_{\text{slow}}^2}{2}\left(\mu r_{\text{slow}}\cos\alpha+\displaystyle\nu r_{\text{fast}}\cos\left(\phi+\alpha\right)\right), \label{eq: three_dynamics2} \\
\end{aligned}\\
\frac{d\phi}{dt} &=
 \begin{aligned}[t]
&\frac{1+r_{\text{fast}}^2}{2}\left(\mu \sin\alpha-\nu \frac{r_{\text{slow}}}{r_{\text{fast}}}\sin\left(\phi-\alpha\right)\right) \\
&-\left(1+r_{\text{slow}}^2\right)\left(\mu \sin\alpha+\nu \frac{r_{\text{fast}}}{r_{\text{slow}}}\sin\left(\phi+\alpha\right)\right). \label{eq: three_dynamics3}
\end{aligned}
\end{align}
From these equations, we found that the reduced system evolves independently of $\Omega$ ,the mean of natural frequency distribution of the oscillators in the slow population.
Later, we will show that this independency holds in general $k$ cases.

\section{Results: Clustered Chimera State}\label{sec:results}

In the previous section, we showed that, if we make a slight modification to Eq. \eqref{eq:original2} and by replacing $\sin\left(\theta^{\text{slow}}_j-\theta^{\text{slow}}_i-\alpha\right)$ with Eq. $\sin\left(2\theta^{\text{slow}}_j-2\theta^{\text{slow}}_i-\alpha\right)$, this modified system with \eqref{eq:original1} in the continuum limit can be simplified to the reduced system (\ref{eq: three_dynamics1}-\ref{eq: three_dynamics3}) by applying the Ott-Antonsen ansatz.
In this section, we numerically simulate the detailed dynamics of the modified and reduced systems.
We set $D=1.0\times10^{-3}$ and imposed the initial condition $r_{\text{fast}}, r_{\text{slow}}\simeq1$ (Each $\theta_i$ was chosen from a uniform distribution in $[0,\pi/30]$), and each population comprised $N=10^4$ oscillators.
The initial order parameters in the reduced systems were set by substituting the conditions of the corresponding $N=10^4$ modified systems.

\begin{figure}[h]
\includegraphics[scale=0.35]{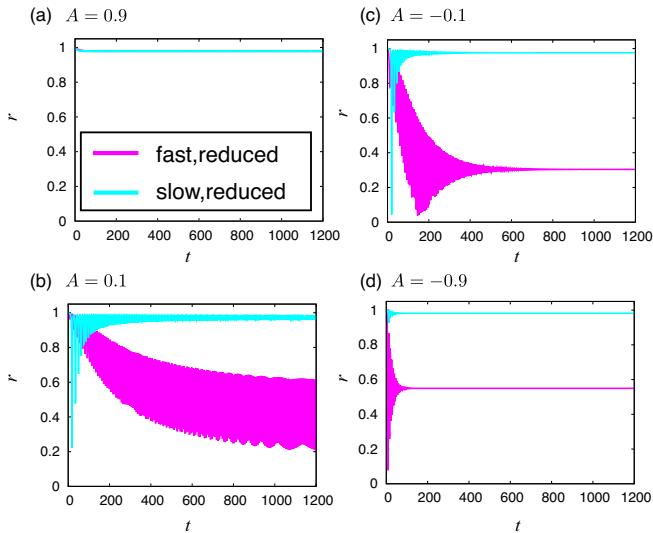}
 \caption{\label{fig: reduced_time_series}Time evolutions of the order parameters of the populations in the reduced system for (a) $A=0.9$, (b) $A=0.1$, (b) $A=-0.1$, and (d) $A=-0.9$.}
\end{figure}

Figure \ref{fig: reduced_time_series} plots the time evolutions of the order parameters of the reduced system for (a) $A=0.9$, (b) $A=0.1$, (c) $A=-0.1$, and (d) $A=-0.9$.
In all cases, we can see the order parameters of the populations reach a steady state.
Panels (a)-(c) reveal three different types of dynamics; coherent, breathing chimera and stable chimera.
In the coherent state (Fig. \ref{fig: reduced_time_series} (a)), the order parameters of both populations approach almost $1$, indicating that the fast and slow oscillators form synchronous clusters.  
In the chimera state, the order parameter of one population becomes incoherent while that of the other population is coherent \cite{abrams2004chimera}.
In this case, the slow oscillators are mutually synchronized while the fast oscillators are not.
In the breathing chimera state (Fig. \ref{fig: reduced_time_series} (b)), the order parameters oscillate.
Conversely, in Fig. \ref{fig: reduced_time_series} (c), the order parameters go finally to the fixed point, which is called the stable chimera state \cite{abrams2008solvable}.
Panels (c) and (d) exhibit similar order parameter behaviors but different phase distributions.
These differences will be thoroughly explored in later simulations of the modified system.

\begin{figure}[h]
\includegraphics[scale=0.35]{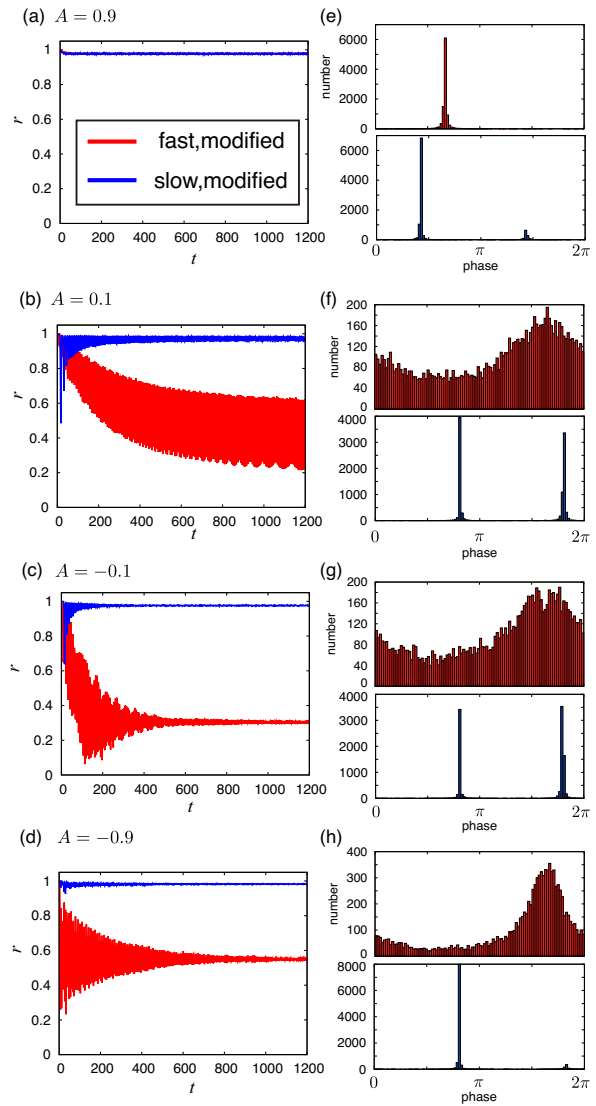}
 \caption{\label{fig: finite_time_series}Time evolutions of the order parameters of the populations in the modified systems; (a) $A=0.9$, (b) $A=0.1$, (b) $A=-0.1$, and (d) $A=-0.9$. Right panels show corresponding snapshots of the steady-state phase distributions (Up: fast population; Down: slow population).}
\end{figure}

Figure \ref{fig: finite_time_series} plots the evolving order parameters and instantaneous phase distributions of the modified system with $N=10^4$.
In panels (a)-(d), the modified systems exhibit the same steady-state behavior as their counterpart reduced systems, but the asymptotic behavior in Fig. \ref{fig: finite_time_series} (d) differs from that Fig. \ref{fig: reduced_time_series} (d).
This difference appears when the coupling between the populations is dominant.

Figure \ref{fig: finite_time_series} also represents snapshots of the steady-state phase distributions of the two populations in the modified system.
For $A=0.1$,  $-0.1$ and $0.9$ (panels (a), (b), and (c), respectively), the fast population is incoherent, whereas the slow population splits into two clusters with a phase difference $\pi$.
Similar states, known as clustered chimera states, have been reported in a delay-coupled system \cite{sethia2008clustered}.

\begin{figure}[h]
\includegraphics[scale=0.7]{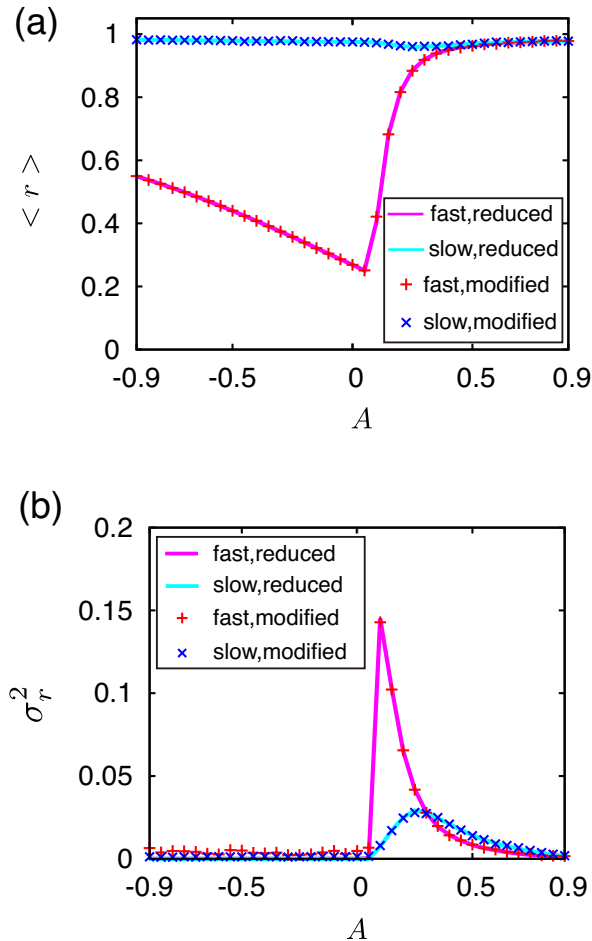}
 \caption{\label{fig ansatz_finite}Comparisons between the average (a) and the standard deviation (b) of the steady-state order parameters in the modified system (with $N=10^4$) and the reduced system.}
\end{figure}

Let us check the result correspondence in the modified and reduced systems.
Panels (a) and (b) of Fig. \ref{fig ansatz_finite}plot the averages and standard deviations, respectively, of the steady-state order parameters in the system reduced by the ansatz and the modified system with $N=10^4$.
When the intra-population coupling strengths are sufficiently strong, both populations settle into coherent states.
Weakened coupling leads to stable or breathing chimera states.

These results were obtained under the initial condition $r_{\text{fast}}, r_{\text{slow}}\simeq1$.
The steady-state order parameters behaved similarly under other starting conditions (data not shown), suggesting that the clustered chimera states in our system are robust.
However, the height ratio of two clusters in the asymptotic phase distributions of the slow oscillators does depend on the initial conditions, because the height difference have no effect on the phase dynamics in Eq. \eqref{eq:modified2}.
Moreover, the transient behaviors crucially depend on the initial conditions in some parameter ranges.

\section{Validity of specific assumption}\label{sec:comparison}

For compatibility with the Ott-Antonsen ansatz, the above analysis imposed an unrealistic assumption on the interactions between the slow oscillators in Eq. \eqref{eq:modified2}.
In real situations, the interaction term between the slow oscillators in Eq. \eqref{eq:original2} should contain first mode in the Fourier series; consequently we need to reassess this assumption.
Unfortunately, without replacing $\sin\left(\theta^{\text{slow}}_j-\theta^{\text{slow}}_i-\alpha\right)$ with $\sin\left(2\theta^{\text{slow}}_j-2\theta^{\text{slow}}_i-\alpha\right)$, we cannot reduce the original system to a low-dimensional system through using the Ott-Antonsen ansatz.
In this section, we investigate whether imposing the above assumption affects the dynamics of two oscillator populations.

To this end, we numerically examine the dynamics of the original system \eqref{eq:original1} and \eqref{eq:original2}.
The natural frequency distributions of the populations are those of the modified system in Secs. \ref{sec:model}-\ref{sec:results}.
As the original system cannot be reduced to a low-dimensional system, we numerically simulate both systems with $N=10^4$.
As in Sec. \ref{sec:results}, we set $D=1.0\times10^{-3}$ and $\alpha=\pi/2-0.05$.
The initial conditions were set to $r_{\text{fast}}, r_{\text{slow}}\simeq1$  (each $\theta_i$ was chosen from uniform distribution in $[0,\pi/30]$).
\begin{figure}[h]
\includegraphics[scale=0.37]{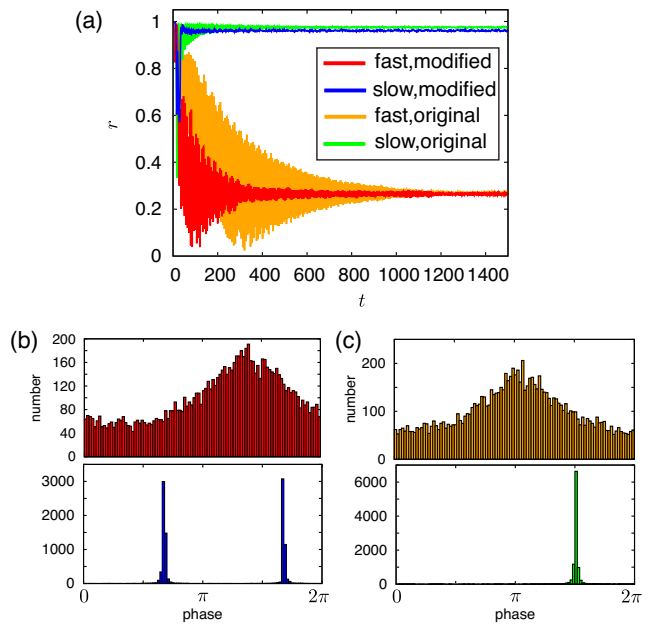}
 \caption{\label{fig:comparison_time_A_0.0}(a) Time evolutions of the order parameters in the modified and original systems with $N=10^4$ for $A=0$.\\Bottom panels are snapshots of the steady-state phase distributions in the modified (b) and original (c) systems for $A=0$.}
\end{figure}
The results for $A=0$ (homogeneous coupling strengths in the system) are plotted in Fig. \ref{fig:comparison_time_A_0.0}.
In Fig. \ref{fig:comparison_time_A_0.0} (a) compares the time evolutions of the order parameters in the modified and original systems with $N=10^4$.
Panels (b) and (c) of this figure are snapshots of the steady-state phase distributions in the modified and original systems, respectively.
In Fig. \ref{fig:comparison_time_A_0.0} (a), the order parameters of both systems reach similar steady states.
However, closer inspection reveals that the asymptotic phase distributions of the slow oscillators differ between the two systems.
Specially, the slow population in the modified system splits into two clusters (Fig. \ref{fig:comparison_time_A_0.0} (b)) but is unimodal in the original system (Fig. \ref{fig:comparison_time_A_0.0} (c)).
Therefore, the original system settles into a normal rather than a clustered chimera state.
The same result emerged under all tested conditions. 
\begin{figure}[h]
\includegraphics[scale=0.7]{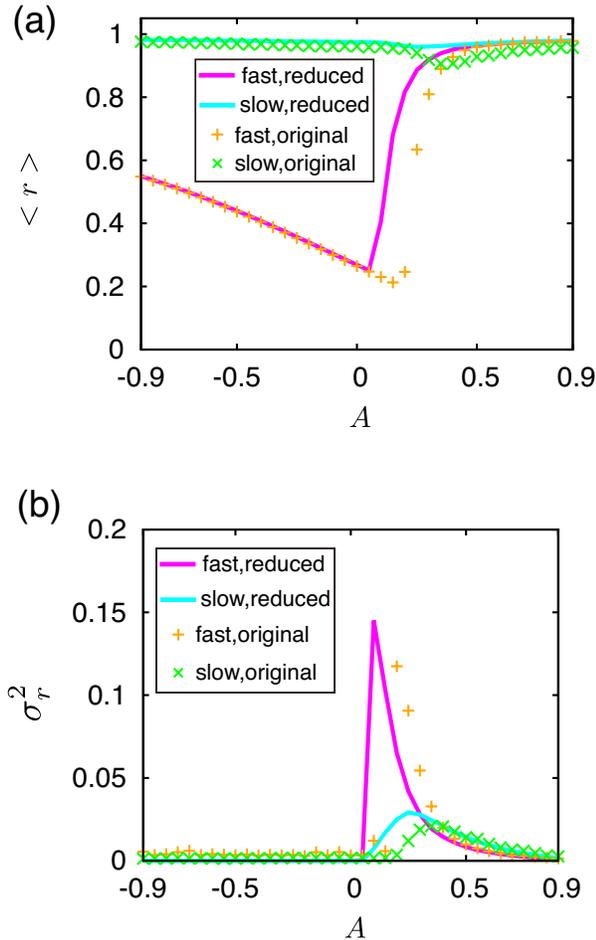}
 \caption{\label{fig:comparison}Comparison between averages (a) and standard deviations (b) of the steady-state order parameters in the reduced and original systems ($N=10^4$).}
\end{figure}
The averages and standard deviations of the steady-state order parameters in the reduced and original systems are compared in Fig. \ref{fig:comparison} (Note that the results of the reduced system are replicated from Fig. \ref{fig ansatz_finite}).
The behaviors of the reduced and original systems are qualitatively similar.
We checked that this similarity remains if $D$ is sufficiently small.

Finally we remark on the different outcomes of the two systems.
In Fig. \ref{fig:comparison}, the asymptotic behaviors of the order parameters differ in certain ranges of the parameter $A$.
The two models differ when $A$ is small and positive; that is, when the intra-population interactions are slightly stronger than the inter-population interactions. 
\begin{figure}[h]
\includegraphics[scale=0.37]{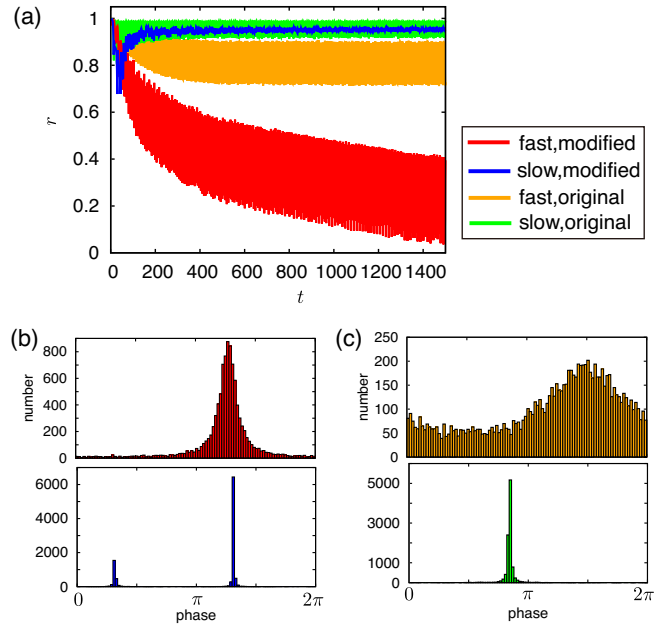}
 \caption{\label{fig:comparison_time_A_0.2}(a) Time evolutions of the order parameters in the modified and original models with $N=10^4$ for $A=0$.\\Bottom panels are snapshots of the steady-state phase distributions in (b) the modified system and (c)the original system for $A=0.2$.}
\end{figure}
To clarify this results, Fig. \ref{fig:comparison_time_A_0.2} (a) plots the evolved order parameter dynamics of the modified and original systems with $N=10^4$ for $A=0.2$.
Fig. \ref{fig:comparison_time_A_0.2} (b) and (c) represent snapshots of the steady-state phase distributions.
In this case, the specific assumption alters the steady-state distribution of the modified system in two ways.
First it amplifies the peak of the fast oscillators relative to the original system.
Second, the slow oscillators separate into two clusters in the modified system, but they form a single cluster in the original system.

\section{general resonant case}\label{sec:general}

As mentioned above, our method is readily extensible from the resonant condition $2:1$ to the general integer resonant condition $k:1$.
Thus, we consider a model of oscillators of two populations under the resonant condition $k:1$:
\begin{align}
\displaystyle\frac{d\theta^{\text{fast}}_i}{dt} &= 
\begin{aligned}[t]
\omega^{\text{fast}}_i +& \displaystyle\frac{\mu}{N}\displaystyle\sum^{N}_{j=1}\sin\left(\theta^{\text{fast}}_j-\theta^{\text{fast}}_i-\alpha\right)\\
&+ \displaystyle\frac{\nu}{N}\displaystyle\sum^{N}_{j=1}\sin\left(k\theta^{\text{slow}}_j-\theta^{\text{fast}}_i-\alpha\right),\label{eq: general1} \\
\end{aligned}\\
\displaystyle\frac{d\theta^{\text{slow}}_i}{dt} &=
\begin{aligned}[t]
\omega^{\text{slow}}_i +& \displaystyle\frac{\mu}{N}\displaystyle\sum^{N}_{j=1}\sin\left(\theta^{\text{slow}}_j-\theta^{\text{slow}}_i-\alpha\right)\\
&+ \displaystyle\frac{\nu}{N}\displaystyle\sum^{N}_{j=1}\sin\left(\theta^{\text{fast}}_j-k\theta^{\text{slow}}_i-\alpha\right), \label{eq: general2}
\end{aligned}
\end{align} 
where the corresponding natural frequency distributions are assumed to obey the Lorentz distributions of Fig. \ref{fig freq_dist}.
Note that the strength parameter $A$ satisfies $\mu=(1+A)/2$ and $\nu=(1-A)/2$.
Thus, by replacing the interaction term $\sin\left(\theta^{\text{slow}}_j-\theta^{\text{slow}}_i-\alpha\right)$ in Eq. \eqref{eq: general2} with $\sin\left(k\theta^{\text{slow}}_j-k\theta^{\text{slow}}_i-\alpha\right)$, we can reduce the original system of Eqs. \eqref{eq: general1} and \eqref{eq: general2} in the continuum limit $N\to\infty$.
After this modification,  the PDF of the slow oscillator population can be expressed as a Fourier series:
\begin{align}
f_{\text{slow}}(\theta,\omega,t) &= \frac{g_{\text{slow}}(\omega)}{2\pi}\left[1+\sum^{\infty}_{m=1}\left(b(\omega,t)^{m}e^{ikm\theta}+\text{c.c.}\right)\right],\notag
\end{align}
where we assume $\left\lvert b\left(\omega,t\right)\right\rvert<1$ under the Ott-Antonsen ansatz.
We also define the complex order parameter for the slow oscillator population $z_{\text{slow}}$:
\begin{align}
\begin{aligned}
\displaystyle z_{\text{slow}}(t)
	&= \int^{\infty}_{-\infty}d\omega\int^{2\pi}_{0}d\theta f_{\text{slow}}\left(\theta,\omega,t\right)e^{ik\theta}.\label{eq:general_slow_order_parameter}
\end{aligned}
\end{align}
Similarly to Sec \ref{sec:reduction}, the population dynamics are finally described by a set of ordinary differential equations with three degrees of freedom:
\begin{align}
\frac{dr_{\text{fast}}}{dt} &=
 \begin{aligned}[t]
&-Dr_{\text{fast}} \\
&+\displaystyle\frac{1-r_{\text{fast}}^2}{2}\left(\mu r_{\text{fast}}\cos\alpha+\displaystyle\nu r_{\text{slow}}\cos\left(\phi-\alpha\right)\right),\label{eq:gen_three_dynamics1} \\
\end{aligned}\\
\frac{dr_{\text{slow}}}{dt} &=
 \begin{aligned}[t]
&-kDr_{\text{slow}} \\
&+\displaystyle\frac{k\left(1-r_{\text{slow}}^2\right)}{2}\left(\mu r_{\text{slow}}\cos\alpha+\displaystyle\nu r_{\text{fast}}\cos\left(\phi+\alpha\right)\right),\label{eq:gen_three_dynamics2} \\
\end{aligned}\\
\frac{d\phi}{dt} &=
 \begin{aligned}[t]
&\frac{1+r_{\text{fast}}^2}{2}\left(\mu \sin\alpha-\nu \frac{r_{\text{slow}}}{r_{\text{fast}}}\sin\left(\phi-\alpha\right)\right) \\
&-\frac{k\left(1+r_{\text{slow}}^2\right)}{2}\left(\mu \sin\alpha+\nu \frac{r_{\text{fast}}}{r_{\text{slow}}}\sin\left(\phi+\alpha\right)\right).\label{eq:gen_three_dynamics3}
\end{aligned}
\end{align}
Here, we used the polar coordinates $z_{\text{fast}}=r_{\text{fast}}e^{-i\phi_{\text{fast}}},z_{\text{slow}}=r_{\text{slow}}e^{-i\phi_{\text{slow}}}$ and denoted the phase difference $\phi=\phi_{\text{fast}}-\phi_{\text{slow}}$ as in Sec. \ref{sec:reduction}. 
We remark that the system dynamics depend only on the ratio of the means of the natural frequencies, $k$.
In other words, the absolute value of the mean frequencies $\Omega$ does not influence the collective behavior of the system.

\begin{figure}[h]
\includegraphics[scale=0.355]{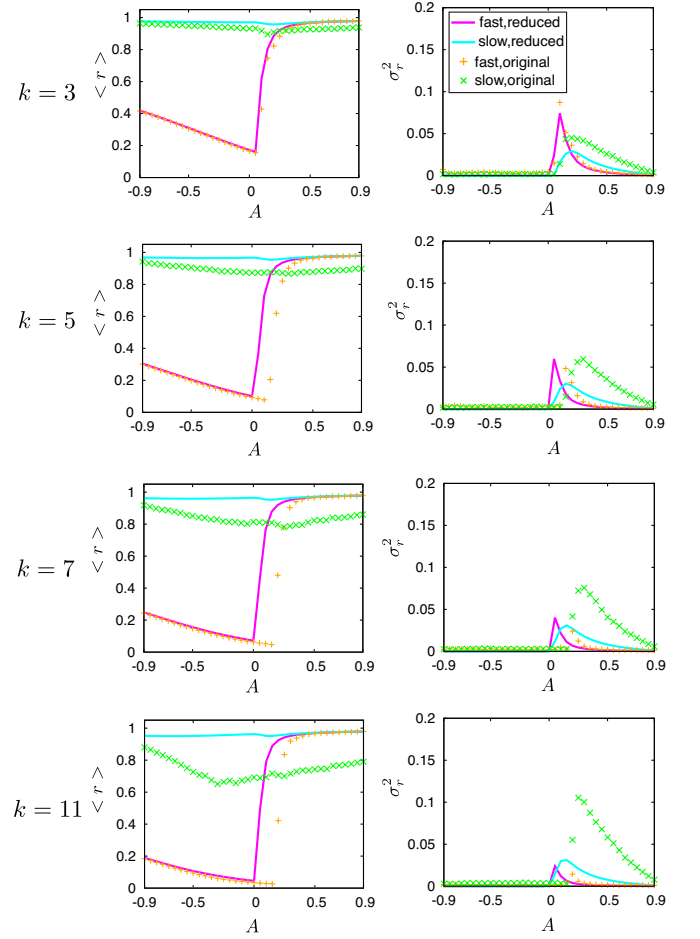}
 \caption{\label{fig:general_k}(Left) Averages and (Right) standard deviations of the steady-state order parameters in the reduced and original systems with $N=10^4$. Top to bottom: $k=3,5,7,11$.}
\end{figure}
Fig. \ref{fig:general_k} shows the averages and standard deviations of the steady-state order parameters in the reduced and modified systems for various $k$ ($3,5,7,11$).
As in Sec. \ref{sec:comparison}, we set $N=10^4$ and the initial conditions $r_{\text{fast}}, r_{\text{slow}}\simeq1$.
We also set $\alpha=\pi/2-0.05$ and $D=1.0\times10^{-3}$.
Qualitatively, the system \eqref{eq:gen_three_dynamics1}-\eqref{eq:gen_three_dynamics3} exhibits the same dynamics as the system \eqref{eq: three_dynamics1}-\eqref{eq: three_dynamics3}, even at higher values of $k$.

However, the asymptotic behaviors differ between the reduced and original systems as $k$ increases.
The widening difference is especially apparent in the standard deviation.
For large $k$, the order parameter of the slow oscillators tends to fluctuate with larger amplitudes in the original system, probably because of the imposed assumption.
In typical Fourier series, the magnitudes of the low-frequency modes are more similar to the first Fourier mode amplitude than those of high-frequency modes.

Under the general condition $k:1$, the modified system develops $k$-clustered chimera states, with $k$ clusters of coherent slow oscillators.
To our knowledge these states have not been previously reported.
The slow oscillators in the $k$-clustered chimera states form $k$ synchronous oscillator groups.
The phase difference between the nearest clusters is approximately $2\pi/k$.
The $k$-clustered chimera states can be  considered as generalized versions of the $2$-cluster ones.
Like the standard chimera state, the $k$-clustered chimera states are classifiable as stable or breathing.

We numerically checked the result correspondence between the reduced and modified systems with large $N$ and general $k$.
Therefore, our proposed reduction is valid in the general $k:1$ case.

\section{Conclusion}\label{sec:discussion}

Referring to phase reduction theory, we investigated the dynamics of multiple populations of phase oscillators.
First, we assumed that the mean of frequency distribution in one population was twice faster than that in the other population.
Applying the Ott-Antonsen ansatz \cite{ott2008low,skardal2011cluster} and imposing an additional assumption, we reduced the original system to a low-dimensional system of ordinary differential equations describing the time evolution of the order parameters.
The population of slow oscillators was treated by the Daido order parameter \cite{daido1992order}.
Clustered chimera states emerged when the inter-population coupling strength was relatively large.

We also investigated the general resonant condition.
Our analysis was extensible from the simple resonant case $2:1$ to the general case $k:1$, where $k$ is any integer.
We confirmed that the result for the case $2:1$ were qualitatively replicated in the general case.
However, for large $k$, our additional assumption significantly altered the dynamics of the original system.

As a future work, we will investigate multifrequency systems completely.
In other words, we can think about dynamics of systems with the more general resonant case $m:n$.
We can apply our approach to the resonant case $m:n$, although it requests us one more assumption on the interactions within populations of fast oscillators.

\begin{acknowledgments}
We thank Takashi Imai and Kaiichiro Ota for fruitful discussions.
This work was supported by Grants-in-Aid from the Ministry of Education, Science, 
Sports, and Culture of Japan: Grant numbers 21120002 and 25115719.
\end{acknowledgments}


\bibliography{res_ref}

\end{document}